\documentclass[12pt,a4paper]{article}
\usepackage[T1]{fontenc}
\usepackage{palatino}
\usepackage{graphics}
\usepackage{setspace}
\makeatletter

\newcommand{\LyX}{L\kern-.1667em\lower.25em\hbox{Y}\kern-.125emX\spacefactor1000}

\usepackage{cite}

\makeatletter\def\diag{\mathop{\operator@font diag}\nolimits}
\makeatletter\def\evmin{\mathop{\operator@font evmin}\nolimits}

\makeatother

\begin{document}

\title{On the reorientation transition of \\
ultra--thin Ni/Cu(001) films }

\author{A.~Hucht and K.~D.~Usadel\\
\emph{\small Theoretische Tieftemperaturphysik, Gerhard-Mercator-Universit\"{a}t,}\\
\emph{\small 47048 Duisburg, Germany}\small }

\date{{}}

\maketitle

\begin{abstract}
The reorientation transition of the magnetization of ferromagnetic films
is studied on a microscopic basis within a Heisenberg spin model. Using
a modified mean field formulation it is possible to calculate properties
of magnetic thin films with non--integer thicknesses. This is especially
important for the reorientation transition in Ni/Cu(001), as there the
magnetic properties are a sensitive function of the film thickness. Detailed
phase diagrams in the thickness--temperature plane are calculated using
experimental parameters and are compared with experimental measurements
by Baberschke and Farle (\textit{J. Appl. Phys.} \textbf{81}, 5038 (1997)).
\end{abstract}
\if0\vspace{2cm}Keywords: reorientation transition, thin films, Heisenberg
model, Ni/Cu(001).
\\
Contact author: 

A. Hucht, 

Theoretische Tieftemperaturphysik, 

Gerhard-Mercator-Universität Duisburg, 

D-47048 Duisburg 

Fon: x49-203-379-3486

Fax: x49-203-379-2965 

Email: fred@thp.Uni-Duisburg.DE

\newpage\renewcommand{\baselinestretch}{1.51}\small\normalsize
\fi

\markright{\rm \emph{Journal of Magn. and Magn. Mater.}, accepted (1998)}

\thispagestyle{myheadings}\pagestyle{myheadings}

The direction of the magnetization of thin ferromagnetic films depends
on various anisotropic energy contributions like surface anisotropy fields,
dipole interaction, and eventually anisotropy fields in the inner layers.
These competing effects may lead to a film thickness and temperature driven
reorientation transition (RT) from an out--of--plane ordered state at low
temperatures to an in--plane ordered state at high temperatures at appropriate
chosen film thicknesses. Experimentally, this transition has been studied
in detail for various ultra--thin magnetic films~\cite{allen1}. Recently,
it was found by Farle \textit{et al.}~\cite{farle1} that ultra--thin Ni--films
grown on Cu(001) show the inverse behavior: the magnetization is oriented
in--plane for thin films and at low temperatures and perpendicular for
thick films and at high temperatures.

It has been shown by various authors~\cite{moschel1,hucht2,jb,hucht3}
that the mechanism responsible for the temperature driven transition can
be understood within the framework of statistical spin models. The RT from
an out--of--plane state at low temperatures to an in--plane state at high
temperature is found to be due to a competition of a positive surface anisotropy
and the dipole interaction. It can occur because the surface anisotropy
has a different temperature dependence that the dipole exchange anisotropy
and vanishes more quickly when approaching the Curie temperature~\cite{jb,hucht3}.
The thickness driven reversed RT in ultra--thin Ni--films is argued to
have its origin in a stress--induced \emph{positive} uniaxial anisotropy
energy in the inner layers with its easy axis perpendicular to the film~\cite{farle1}.
This anisotropy is in competition with the dipole interaction and with
a \emph{negative} surface anisotropy. This scenario can indeed lead to
a reversed \emph{temperature driven} RT, but now the reduced surface magnetization
plays a crucial role~\cite{jb,hucht3}. A third type of RT has recently
been found theoretically, where the magnetization switches from perpendicular
to in--plane direction with \emph{increasing} temperature, but with \emph{decreasing}
film thickness~\cite{hucht4}.

In this paper we will focus on the phase diagram of Ni/Cu(001) and on the
influence of microscopic fourth--order anisotropy terms. The calculations
are done in the framework of a classical ferromagnetic Heisenberg model
consisting of \( L \) two--dimensional layers on a face centered cubic
(001) lattice. The Hamiltonian reads 
\begin{eqnarray}
\mathcal{H} & = & -\frac{J}{2}\sum _{\langle ij\rangle }\vec{s}_{i}\cdot \vec{s}_{j}-\sum _{i}D_{2}^{i}(s_{i}^{z})^{2}+D_{4\perp }^{i}(s_{i}^{z})^{4}\nonumber \\
 &  & +\frac{\omega }{2}\sum _{ij}r_{ij}^{-3}\vec{s}_{i}\cdot \vec{s}_{j}-3r_{ij}^{-5}(\vec{s}_{i}\cdot \vec{r}_{ij})(\vec{r}_{ij}\cdot \vec{s}_{j}),\label{hami} 
\end{eqnarray}
 where \( \vec{s}_{i}=(s_{i}^{x},s_{i}^{y},s_{i}^{z}) \) are spin vectors
of unit length at position \( \vec{r}_{i}=(r_{i}^{x},r_{i}^{y},r_{i}^{z}) \),
and \( \vec{r}_{ij}=\vec{r}_{i}-\vec{r}_{j} \). \( J \) is the nearest-neighbor
exchange coupling constant. The uniaxial and fourth--order anisotropies
\( D_{n}^{i} \) are position--dependent: \( D_{n}^{i}=D_{n}^{\mathrm{S}} \)
if spin \( i \) is in the top layer, \( D_{n}^{i}=D_{n}^{\mathrm{S'}} \)
if spin \( i \) is in the bottom layer, and \( D_{n}^{i}=D_{n}^{\mathrm{V}} \)
otherwise. Finally, \( \omega =\mu _{0}\mu ^{2}/4\pi a^{3} \) is the strength
of the long range dipole interaction on a lattice with next--neighbor distance
\( a \) (\( \mu _{0} \) is the magnetic permeability and \( \mu  \)
is the effective magnetic moment of one spin).

The Hamiltonian Eq.~(\ref{hami}) is handled in a molecular-field approximation~\cite{hucht3}.
In the following we assume translational invariance within the layers and
furthermore that the magnetization is homogeneous inside the film and only
deviates at the surfaces~\cite{hucht4}. Therefore we can set \( \langle \vec{s}_{i}\rangle =\langle \vec{s}_{\lambda }\rangle  \)
if \( \vec{s}_{i} \) is a spin in a volume layer (\( \lambda =\mathrm{V} \))
or in the top or bottom surface layer (\( \lambda =\mathrm{S},\mathrm{S'} \)).
We will focus on the case \( L>2 \), i.e. both surfaces are complete.
The resulting system with three mean field spins has the effective interactions
\( \tilde{x}_{\lambda \mu } \), where \( x_{\delta } \) is either the
number of next neighbors, \( z_{\delta } \), or dipole sum in fcc(001)
geometry, \( \Phi _{\delta }=\sum _{i,j}'\frac{i^{2}+j^{2}+(i+j)\delta -\delta ^{2}/2}{(i^{2}+j^{2}+(i+j)\delta +\delta ^{2})^{5/2}} \),
respectively, between layers \( \lambda  \) and \( \lambda +\delta  \).
We have \( \tilde{x}_{\mathrm{SS'}}=\tilde{x}_{\mathrm{S'S}}=0 \), \( \tilde{x}_{\mathrm{SS}}=\tilde{x}_{\mathrm{S'S'}}=x_{0} \),
\( \tilde{x}_{\mathrm{SV}}=\tilde{x}_{\mathrm{S'V}}=x_{1} \), \( \tilde{x}_{\mathrm{VV}}=x_{0}+2x_{1}(1-L_{\mathrm{V}}^{-1}) \),
and \( \tilde{x}_{\mathrm{VS}}=\tilde{x}_{\mathrm{VS'}}=x_{1}L_{\mathrm{V}}^{-1} \),
with \( L_{\mathrm{S}}=L_{\mathrm{S'}}=1 \) and \( L_{\mathrm{V}}=L-2 \).
The constants \( z_{\delta } \) and \( \Phi _{\delta } \) are \( z_{0}=z_{1}=4 \),
\( \Phi _{0}=9.034 \) and \( \Phi _{1}=1.429 \). \( \Phi _{\delta >1} \)
can be neglected in our approach. Note that the coupling between the volume
spin and the surface spins is asymmetric. With this method we are not restricted
to films with integer values of the thickness anymore. The effective interactions
\( \tilde{x}_{\lambda \mu } \) enter the mean field Hamiltonian via the
mean fields in layer \( \lambda  \), \( \vec{h}_{\lambda }=\sum _{\mu =1}^{3}J\tilde{z}_{\lambda \mu }\vec{m}_{\mu }+\tilde{\Phi }_{\lambda \mu }\mathbf{W}\vec{m}_{\mu } \),
with the dipole exchange anisotropy \( \mathbf{W}=\diag (1/2,1/2,-1) \).
In order to determine the phase diagram of this Hamiltonian in the thickness--temperature
plane, we directly calculate the phase boundaries using a stability analysis
of the mean field free energy. Defining the function \( e_{\textrm{min}}=\evmin (\mathbf{A}) \)
which returns the smallest eigenvalue \( e_{\textrm{min}} \) of the matrix
\( \mathbf{A} \), we can calculate the effective anisotropy \( K_{2} \)
from the Hessian of \( \mathcal{F}^{\mathrm{MF}} \) with respect to the
azimuth angle \( \vartheta _{\lambda } \) of the magnetization \( \vec{m}_{\lambda } \)~\cite{note1}
\begin{equation}
K_{2}(L,T,\vartheta )=\frac{1}{2}\evmin \left( \left. \frac{\partial ^{2}\mathcal{F}^{\mathrm{MF}}(L,T)}{\partial \vartheta _{\lambda }\partial \vartheta _{\mu }}\right| _{\vartheta _{\nu }=\vartheta }\right) .
\end{equation}
 Now consider the phase with magnetization parallel to the film normal
\( \vec{z} \), where \( \vartheta _{\lambda }=0 \). At thicknesses \( L \)
and temperatures \( T \) where this phase is stable, we have \( K_{2}(L,T,0)>0 \).
At the phase boundaries both to the canted phase and to the paramagnetic
phase, \( K_{2}(L,T,0) \) becomes zero, while it is negative in the canted
phase and in the phase with in--plane magnetization. On the other hand,
\( K_{2}(L,T,\pi /2) \) is positive when the in--plane phase is stable
and becomes negative in the phases with perpendicular component of the
magnetization. In the paramagnetic phase \( K_{2}(L,T,\vartheta )=0 \)
for all \( \vartheta  \). Hence the reorientation temperatures are given
by \( K_{2}(L,T_{r}^{xy}(L),0)=0 \) and \( K_{2}(L,T_{r}^{z}(L),\frac{\pi }{2})=0 \).
Next we will examine the phase diagram of this model for parameters measured
on Ni/Cu(001)~\cite{bab1}: The exchange interaction is approximated via
the Curie temperature \( T_{c}=631\, \textrm{K} \) of bulk Ni, using the
mean field formula \( 3T_{c}=zJ \) with \( z_{\mathrm{f}cc}=12 \) for
the isotropic classical Heisenberg model to give \( J\simeq 13.6\, \textrm{meV} \).
This is a rather rough estimate, but note that the exact value of \( J \)
has nearly no influence on the following results, as long as \( J \) is
large compared to the other energies in the model. With \( \mu =0.62\mu _{\mathrm{B}} \)
and \( a=2.5 \)\AA ~we get \( \omega =1.3\, \mu \textrm{eV }\simeq 10^{-4}J \)
for the dipole constant. The uniaxial surface anisotropy is \( (D_{2}^{\mathrm{S}}+D_{2}^{\mathrm{S'}})/2=-60\, \mu \textrm{eV }\simeq -47\omega  \)
and the uniaxial volume anisotropy is \( D_{2}^{\mathrm{V}}=+40\, \mu \textrm{eV }\simeq 31\omega  \).
Note that our surface anisotropies are the sum of the experimental surface
and volume part, as our surface layers also count to the volume. For simplicity
we assume that the bottom surface carries the volume anisotropies, hence
we get \( D_{2}^{\mathrm{S}}\simeq -125\omega  \).

\begin{figure}
{\centering \resizebox*{14cm}{!}{\includegraphics{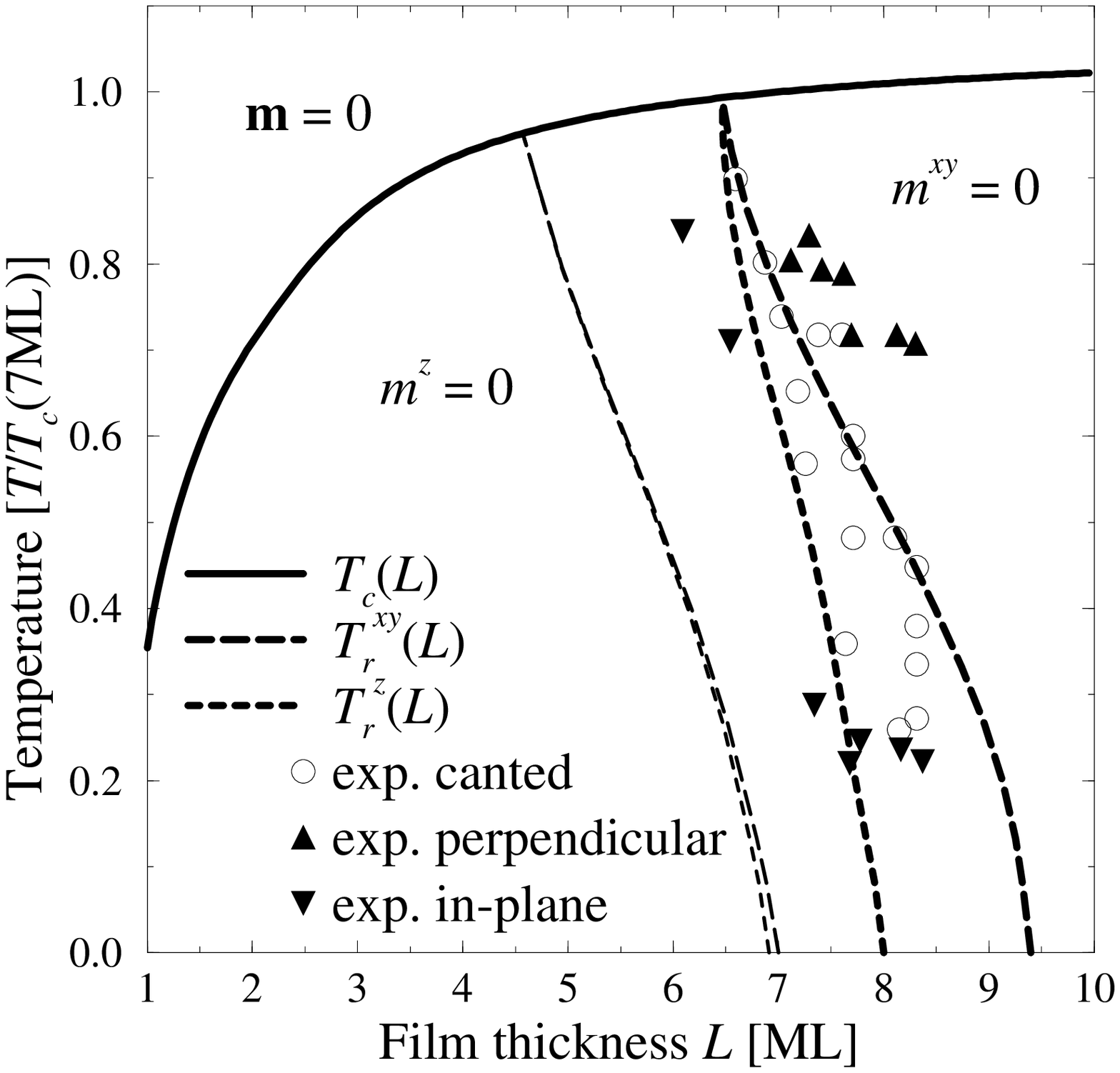}} \par}

\caption{\label{fig}Calculated phase diagram for Ni/Cu(001). For the thick lines,
the model parameters are \protect\( \omega =10^{-4}J\protect \), \protect\( D_{2}^{\mathrm{S}}=-100\omega \protect \),
\protect\( D_{2}^{\mathrm{V}}=D_{2}^{\mathrm{S}'}=24\omega \protect \),
\protect\( D_{4\perp }^{\mathrm{S}}=24\omega \protect \), \protect\( D_{4\perp }^{\mathrm{V}}=D_{4\perp }^{\mathrm{S}'}=-4\omega \protect \).
The thick solid line represents the Curie temperature \protect\( T_{c}(L)\protect \)
of the film, the dotted and dashed lines are \protect\( T_{r}^{z}(L)\protect \)
and \protect\( T_{r}^{xy}(L)\protect \), respectively. The symbols are
experimental measurements of the reorientation transition in Ni/Cu(001)
taken from~\cite{bab2}. The temperature axis is normalized to the Curie
temperature of a film with thickness \protect\( L=7\, \textrm{ML}\protect \).
The parameters for the thin lines are described in the text. }
\end{figure}

To check the influence of \( D_{4\perp }^{\lambda } \), we show the phase
diagram for the abovementioned parameter set with \( D_{4\perp }^{\lambda }=0 \).
The phase boundaries are the thin lines in figure~\ref{fig}. We obtain
a reversed RT as expected for Ni/Cu(001), the reorientation thickness ranges
from 5ML to 7ML which is slightly lower as in the experiment~\cite{bab2}.
We find that the canted phase exist, but it is rather narrow at these parameters.
This canted phase is stable because we allow non--collinear magnetizations
in different layers. To reproduce the experimental finding of the rather
broad canted phase, we set the fourth--order anisotropies \( D_{4\perp }^{\lambda }\simeq -\frac{1}{4}D_{2}^{\lambda } \)~\cite{bab2}.
The phase diagram of our model with these values and slightly modified
\( D_{2}^{\lambda } \) is depicted in figure~\ref{fig} together with
experimental data obtained by Baberschke and Farle~\cite{bab2}. As the
mean field theory always overestimates \( T_{c} \), especially for low
dimensional systems, the critical line \( T_{c}(L) \) from~\cite{bab2}
cannot be reproduced well within our theory. Thus we normalized the temperature
axis not with \( T_{c}^{\mathrm{bulk}} \), but with the Curie temperature
of a film with \( L=7\, \textrm{ML} \).

In this paper we defined a modified mean field model to describe the RT
in Ni/Cu(001). Using appropriate model parameters close to experimental
findings, we find a rather nice agreement of theory and experiment. To
reproduce the width of the phase with canted magnetization, we need to
introduce microscopic fourth--order anisotropies \( D_{4\perp } \).

This work was supported by the Deutsche Forschungsgemeinschaft through
Sonderforschungsbereich 166. Discussion with K. Baberschke is gratefully
acknowledged.

\newpage

\end{document}